\documentclass[]{article}

\pdfoutput=1

\usepackage[authoryear]{natbib}
\usepackage{booktabs}  
\usepackage{amsfonts}
\usepackage{amsmath}
\usepackage{dcolumn}
\usepackage{bm}
\usepackage{cancel}
\usepackage{verbatim}
\usepackage{ifthen}
\usepackage{url}
\usepackage{sectsty}
\usepackage{balance}
\usepackage{graphicx} 
\usepackage{lastpage}
\usepackage[format=plain,justification=RaggedRight,singlelinecheck=false,font=small,labelfont=bf,labelsep=space]{caption}
\usepackage{fancyhdr}
\usepackage{subcaption}
\usepackage{makecell}
\usepackage{multirow}
\usepackage{fancyvrb}
\usepackage[ruled,vlined]{algorithm2e}
\usepackage{authblk}
\usepackage[letterpaper, margin=1in]{geometry} 

\usepackage{setspace}
\usepackage{cellspace}
\newcommand{\Real}{\mathbb{R}}

\newcommand{\Tra}{^{\sf T}} 
\newcommand{\Inv}{^{-1}} 
 
\newcommand{\V}[1]{{\bm{\mathbf{\MakeLowercase{#1}}}}} 
\newcommand{\M}[1]{{\bm{\mathbf{\MakeUppercase{#1}}}}} 

\def\b{{\bf b}}

\def\E{{\bf E}} 
\def\e{{\bf e}}

\def\I{{\bf I}}

\def\t{{\bf t}}

\def\X{{\bf X}} 
\def\x{{\bf x}} 

\def\XtX{{\X\Tra \X}} 

\def\Y{{\bf Y}}
\def\y{{\bf y}}

\def\z{{\bf z}}

\def \valpha{{\V{\alpha}}}

\def \vmu{{\V{\mu}}}

\def \vdelta{{\V{\delta}}}

\def \msigma{{\M{\Sigma}}}

\def \vtheta{{\V{\theta}}}

\newcommand{\normal}{\mathcal{N}}

\newcommand{\zerovec}{\V{0}}

\newcommand{\Eq}{\mathbb{E}_{q}}

\DeclareMathOperator*{\argmax}{arg\,max}
\DeclareMathOperator*{\argmin}{arg\,min}
\let\code=\texttt
\newcommand{\fct}[1]{\code{#1()}}

\DefineVerbatimEnvironment{Sinput}{Verbatim}{fontshape=sl}
\DefineVerbatimEnvironment{Soutput}{Verbatim}{}
\DefineVerbatimEnvironment{Scode}{Verbatim}{fontshape=sl}

\DefineVerbatimEnvironment{Code}{Verbatim}{}
\DefineVerbatimEnvironment{CodeInput}{Verbatim}{fontshape=sl}
\DefineVerbatimEnvironment{CodeOutput}{Verbatim}{}
\newenvironment{CodeChunk}{}{}

\let\proglang=\textsf
\newcommand{\pkg}[1]{{\fontseries{b}\selectfont #1}}
\newcommand{\bigcell}[2]{\begin{tabular}{@{}#1@{}}#2\end{tabular}}

\providecommand{\keywords}[1]{\textbf{\textit{Keywords:}} #1}

\title{Variational Inference for Shrinkage Priors: The \proglang{R} package 
\pkg{vir}}
\author[1]{Suchit Mehrotra}
\author[1]{Arnab Maity}
\affil[1]{Department of Statistics, North Carolina State University}
\date{}

\begin{document}
	
\maketitle

\doublespacing

\begin{abstract} 
\noindent We present \pkg{vir}, an \proglang{R} package for variational 
inference with
shrinkage priors. Our package implements variational and stochastic variational 
algorithms for linear and probit regression models, the use of which is a 
common first step in many applied analyses. We review variational inference and 
show how the derivation for a Gibbs sampler can be easily modified to derive a 
corresponding variational or stochastic variational algorithm. We provide 
simulations showing that, at least for a normal linear model, variational inference 
can 
lead to similar uncertainty quantification as the corresponding Gibbs samplers, 
while estimating the model parameters at a fraction of the computational cost. Our 
timing experiments show situations in which our algorithms converge faster than 
the frequentist LASSO implementations in \pkg{glmnet} while simultaneously 
providing superior parameter estimation and variable selection. 
Hence, our package can be utilized to quickly explore different combinations of 
predictors in a linear model, while providing accurate uncertainty quantification in 
many applied situations.
The package is implemented natively in
\proglang{R} and \pkg{RcppEigen}, which has the benefit of bypassing the
substantial operating system specific overhead of linking external libraries to
work efficiently with \proglang{R}.
\end{abstract}

\keywords{Bayesian statistics, big data, Gibbs sampling, shrinkage priors, 
variational inference}

\section{Introduction} \label{sec:vir:intro}

Bayesian statistics assumes that all information about a parameter of interest
is contained in a probability distribution called the posterior. In most modern
problems, the posterior is hard to compute due to an intractable normalizing
constant, and approximations to it are necessary to conduct inference. The most
popular choice for approximating the posterior distribution has been the use of
Markov Chain Monte Carlo (MCMC) algorithms \citep{robert2013monte}.  MCMC
algorithms work by constructing a Markov chain which has the posterior as its
stationary distribution. Once the stationary distribution is reached, samples
collected from chain serve as an approximation of the posterior distribution. 

A variety of MCMC algorithms exist, with the two most important being the
Metropolis-Hastings algorithm \citep{metropolis1953, hastings1970monte} and the
Gibbs sampler \citep{geman1984stochastic, gelfand1990sampling}. The use of these
algorithms is simplified by software which implement them in an automatic
fashion by allowing the user to define a model; the \proglang{R} packages
\pkg{rjags} \citep{plummer2019rjags} and \pkg{r2winbugs} \citep{r2winbugs}
implement Gibbs samplers, while the package \pkg{rstan} \citep{stan2018rstan}
uses the No-U-Turn sampler \citep{hoffman2014no} for Hamiltonian Monte Carlo
\citep{betancourt2017conceptual}.

Unfortunately, an issue with MCMC based approaches is that they scale poorly
with dataset size or when the number of parameters is large. Consequently,
modern work in MCMC has focused on dealing with these limitations. To name a
few, these methods work by exploiting modern computing architecture such as GPUs
\citep{terenin2019gpu}, splitting the dataset into smaller chunks (data
sharding) and running independent chains asynchronously
\citep{terenin2020asynchronous} or combining the results after convergence
\citep{consensusmc, srivastava2015wasp}, compressing the data before analysis
\citep{bayesiancompressed}, sub-sampling data to approximate expensive
likelihoods \citep{quiroz2018speeding, sgdmcmc}, or using low-rank proposals for
high dimensional parameter spaces \citep{saibaba2019efficient}.  Most of these
approaches are not readily available for use in statistical software, something
which inhibits the use of the Bayesian paradigm in many applied problems of
interest. 

An alternative to sampling based approaches for posterior approximation is to
use variational inference (VI) \citep{blei2017variational}, which casts the
problem into an optimization framework. The main idea is to find an optimal
distribution from a family of densities which is closest to the posterior based
on some distance measure, where the family is chosen to balance computational
tractability and the quality of the posterior approximation. Since variational
inference is an optimization based approach, it has the advantage that it can be
easily scaled to large data sets using stochastic optimization
\citep{hoffman2013svi}.  At present, only a few \proglang{R} packages exist
which implement variational methods: \pkg{rstan}, which utilizes the \pkg{STAN}
framework to implement automatic differentiation variational inference (ADVI),
but includes a warning that their implementation is unstable and subject to
change \citep{stan2018rstan, advipaper}, and the package \pkg{varbvs}
\citep{carbonetto2012scalable}, which incorporates methods for variable
selection using spike-and-slab priors for normal and logistic linear models.

In this chapter we present a new \proglang{R} package \pkg{vir}, which includes
a set of variational and stochastic variational algorithms to estimate
parameters in linear and probit regression with shrinkage priors.  We
incorporate the normal (ridge), double-exponential (LASSO)
\citep{park2008lasso}, and horseshoe \citep{carvalho2010horseshoe} priors to
conduct variable selection and inference, a problem which arises as a first step
in almost all applied analyses.  Our package adds to the \proglang{R} ecosystem
by providing a suite of computationally efficient variational algorithms, which
scale with both the number of parameters, by allowing independence assumptions
between regression coefficients, and the number of data points, by utilizing
stochastic optimization methods. We implement the algorithms natively in
\proglang{R} and \pkg{RcppEigen}, which has the benefit of bypassing the
substantial operating system specific overhead of linking external libraries to
work efficiently with \proglang{R}.  Through our simulation studies, we show
that the variational algorithms presented in this chapter are competitive with the
popular \pkg{glmnet} package \citep{glmnet}, which is widely used for variable
selection, in both computation time and variable selection accuracy.
Additionally, our simulation studies calculate empirical coverage probabilities
for the regression coefficients, showing that the variational algorithms have
the potential to recover the correct coverage in many applied scenarios. 

The rest of this chapter proceeds as follows: Section \ref{sec:vir:mcmc}
provides a short review of Gibbs sampling, Section \ref{sec:vir:vi} reviews
relevant details for variational and stochastic variational inference, Section
\ref{sec:vir:usage} reviews the use and implementation of our package, Sections
\ref{sec:vir:simulations} and \ref{sec:vir:timing} contain numerical studies
comparing \pkg{vir} with \pkg{glmnet}, and Section \ref{sec:vir:discussion}
provides a short discussion of our results. 

\section{Markov Chain Monte Carlo Methods} \label{sec:vir:mcmc}

As discussed in Section \ref{sec:vir:intro},  the core problem in conducting
Bayesian inference for a parameter, $\vtheta$, is the calculation of its
posterior distribution conditioned on the data, $\y$: $p(\vtheta | \y)$.  This
quantity is rarely available in closed form, and approximations to it are
necessary to estimate functions of $\vtheta$ that may be of interest.  The
variational algorithms implemented in this package are an example of such
approximation algorithms, but their use in the Bayesian literature pales in
comparison to the use of Markov Chain Monte Carlo (MCMC) methods. In this
section, we focus our discussion of MCMC on Gibbs sampling because of its close
relationship with variational inference: the derivations necessary to derive and
implement a Gibbs sampler can be extended to derive the corresponding
variational algorithms. For a comprehensive treatment of MCMC algorithms, we
refer the reader to \citet{robert2013monte} and \citet{brooks2011handbook}.

\subsection{Gibbs Sampling}

Gibbs sampling is one of the most popular MCMC algorithms in use today. It is a
widely applicable special case of the Metropolis-Hastings algorithm and is
straightforward to use when the full conditional distributions of each parameter
are easy to sample from.

To understand how these algorithms are implemented, first note that the
posterior distribution, $p(\vtheta | \y)$ can also be written as $p(\theta_1,
\dots, \theta_P | \y)$. We then choose an initial state of the Markov chain for
$\vtheta$, $(\theta_1^{(0)}, \dots, \theta_P^{(0)})$ and update this state one
element at a time by sampling from its full conditional, the distribution of
that parameter conditioned on all others. If we continue this procedure for many
iterations, updating each element one at a time in no particular order, the
Markov chain will converge to the posterior distribution of interest, and
storing the state at the end of each iteration after a `burn-in' period will
give us approximate draws from $p(\vtheta | \y)$; the full description of the
algorithm is given in Algorithm \ref{algo:vir:gibbs-sampler}. An important
technique to allow the use of a Gibbs sampler in a model is to use conjugate
priors for the parameters, which lead to each full conditional being of the same
form as the prior distribution. All the models we consider in this chapter have
priors with this conjugate relationship.  

\begin{algorithm}[t]
	\SetAlgoLined
	\KwResult{Samples from a posterior approximation}
	\KwIn{Integers: \code{burn\_in} and \code{n\_iter}, with  $\code{burn\_in} < 
	\code{n\_iter}$}
	Set $\vtheta^{(0)} = (\theta_1^{(0)}, \dots, \theta_P^{(0)})$ \;
	\For{$i\gets0$ \KwTo \code{n\_iter}}{
		$\theta_1^{(i+1)} \sim 
			p(\theta_1| \theta_2^{(i)}, \theta_3^{(i)}, \dots, \theta_P^{(i)})$\;
		$\theta_2^{(i+1)} \sim 
			p(\theta_2| \theta_1^{(i+1)}, \theta_3^{(i)}, \dots, \theta_P^{(i)})$\;
		\vdots
		$\theta_p^{(i+1)} \sim 
			p(\theta_p| \theta_1^{(i+1)}, \theta_2^{(i+1)}, \dots, 
			\theta_{p-1}^{(i+1)}, \theta_{p+1}^{(i)}, \dots, \theta_P^{(i)})$\;
		\vdots
		$\theta_P^{(i+1)} \sim 
			p(\theta_p| \theta_1^{(i+1)}, \dots \theta_{P-1}^{(i+1)})$\;		
	}
	\KwOut{$\left\{\vtheta^{(\code{burn\_in} + 1)}, \dots, \vtheta^{(\code{n\_iter})} 
	\right\}$}
	\caption{Gibbs Sampler} \label{algo:vir:gibbs-sampler}
\end{algorithm}

\subsubsection{Exponential Families}

Finding a conjugate prior for a distribution is always possible for regular
exponential families (\citet{bernardo2000}, Proposition 5.4), where the
probability density (mass) function for a random variable, $\y$, can be written
in the form: 
\begin{align} \label{eq:vir:exp-fam}
	f(\y | \vtheta) = h(\y) \exp \left\{ 
		\langle \vdelta_y(\vtheta) | \t(\y) \rangle - 
		A[\vdelta_y(\vtheta)]
	\right\},
\end{align}
where $\vdelta_y(\vtheta)$ is called the natural parameter of the distribution,
$\t(\y)$ is a vector of sufficient statistics, and $\langle \cdot | \cdot
\rangle$ is an inner product. It should be noted that $\t(\y)$ lies in a general
vector space and its elements can be scalars, vectors, or matrices, with the
inner product $\langle \cdot | \cdot \rangle$ defined to be the sum of the inner
products of the respective spaces. For example, a $p$ dimensional multivariate
normal distribution for a random variable $\y$, has sufficient statistics $\y
\in \Real^p$ and $\y\y\Tra \in \Real^{p \times p}$. Hence, $\t(\y) \in
\mathbb{R}^p \times \Real^{p \times p}$ with $\langle \cdot | \cdot \rangle =
\langle \cdot | \cdot \rangle_{\Real^p} + \langle \cdot | \cdot
\rangle_{\Real^{p \times p}}$.  

For any distribution that can be written in the form of \eqref{eq:vir:exp-fam},
there exists a conjugate prior for the parameter $\vtheta$ that can be written
in the same form: 
\begin{align*}
	p(\vtheta | \valpha) = h(\vtheta) \exp \left\{ 
		\langle \valpha | \t(\vtheta) \rangle - 
		A(\valpha)	\right\},
\end{align*}
where the sufficient statistic $\t(\vtheta)$ is, 
\begin{align*}
	\t(\vtheta) = 
	\begin{pmatrix}
		\vdelta_{y}(\vtheta) \\
		-A[\vdelta_y(\vtheta)]
	\end{pmatrix},
\end{align*} 
and $\valpha$, the natural parameter of the prior distribution, can be
partitioned into $\valpha = (\valpha_1, \alpha_2)^T$, with $\valpha_1$ being of
the same dimension as $\vdelta_{y}(\vtheta)$ and $\alpha_2$ being a scalar.
Assuming we have $N$ independent and identically distributed samples from $f(\y
| \vtheta)$, multiplying the likelihood with the prior we see that:
\begin{align*}
	p(\vtheta | \y_1, \dots, \y_N, \valpha) & \propto 
		\prod_{n = 1}^N p(\y_n | \vtheta) p(\vtheta | \valpha), \\
	& \propto \prod_{n = 1}^N \exp \left\{ 
		\langle \vdelta_y(\vtheta) | \t(\y_n) \rangle - A[\vdelta_y(\vtheta)]
	\right\}
	h(\vtheta) \exp \left\{ 
		\langle \valpha| \t(\vtheta) \rangle - 
		A(\valpha)
	\right\}, \\
	& \propto h(\vtheta) \exp \left \{
		\langle \vdelta_{\vtheta}^* | \t(\vtheta) \rangle
		- A(\valpha)
	\right \},
\end{align*}
where, 
\begin{align}
\vdelta^*_{\vtheta} = 
	\begin{pmatrix}
		\sum_{n = 1}^N \t(\y_n) + \valpha_1 \\
		N + \alpha_2
	\end{pmatrix},
\end{align} 
and we use the fact that $-A[\vdelta_y(\vtheta)] = \langle
-A[\vdelta_y(\vtheta)] | 1 \rangle_{\Real}$.

Such manipulations with exponential families are of critical importance when
deriving variational algorithms. We will see that writing the full conditionals
of a Gibbs sampler in a natural parameter exponential family form simplifies the
derivation of the stochastic gradient descent algorithms implemented in the
package. 

\section{Variational Inference} \label{sec:vir:vi}

MCMC methods are known to scale poorly to large datasets, both in the number of
observations ($N$) and the number of parameters ($P$).  Variational Inference
\citep{vbreview, bishop2006pattern, murphy2012} aims to alleviate these issues
by approximating the probability distribution of interest by utilizing
optimization instead of sampling. As will be seen in our numerical experiments
in Section \ref{sec:vir:simulations}, when compared with Gibbs sampling, these
methods yield similar results for many quantities of interest while
approximating the posterior at a fraction of the computational cost. In the
following sections we will review some of the salient details of variational
inference while referring the reader to \citet{vbreview, hoffman2013svi,
bishop2006pattern, murphy2012} for a thorough review.

\subsection{General Setup}

For the rest of this section, let $p( \vtheta | \Y)$, be the posterior
distribution of interest, with $\vtheta$ being the parameter and $\Y$ being the
observed data, $\Y = (\y_1, \dots, \y_n)$. While MCMC aims to sample from this
distribution, VI aims to approximate it minimizing the Kullbak-Leibler (KL)
divergence using a family of candidate densities, $\mathcal{D}$. The
optimization problem is: 
\begin{align} 
\begin{split} \label{eq:vi_optim}
	q^*(\vtheta) & = \argmin_{q(\vtheta) \in \mathcal{D}} 
		KL \left\{q(\vtheta), \ p(\vtheta | \Y) \right\}, \\
	& = \argmin_{q(\vtheta) \in \mathcal{D}} 
		\int q(\vtheta) \log \left\{ \frac{q(\vtheta)}{p(\vtheta | \Y)} \right\} d\vtheta.
\end{split}
\end{align}
By utlizing Bayes' rule, we can write
\eqref{eq:vi_optim} as a maximization problem:
\begin{align}
\begin{split}
\label{eq:elbo}
\argmin_{q(\vtheta) \in \mathcal{D}} KL \left \{ q(\vtheta), \ p(\vtheta | \Y) 
\right\} & = 
\argmin_{q(\vtheta) \in \mathcal{D}}  \left\{ 
\Eq[\log q(\vtheta)] - \Eq[\log p(\vtheta | \Y)] 
\right\}, \\
& =  \argmin_{q(\vtheta) \in \mathcal{D}} \left\{ 
\Eq[\log q(\vtheta)] - \Eq[\log p(\vtheta, \Y)] + \log p(\Y)
\right	\}, \\
& = \argmin_{q(\vtheta) \in \mathcal{D}}  \left\{ 
\Eq[\log q(\vtheta)] - \Eq[\log p(\vtheta, \Y)]
\right\},  \\
& = \argmax_{q(\vtheta) \in \mathcal{D}}  \left\{ 
\Eq[\log p(\vtheta, \Y)] - \Eq[\log q(\vtheta)] 
\right\},
\end{split}
\end{align}
where the second to last line drops $\log p(\Y)$ because it does not depend on
$\vtheta$ and the last multiplies the equation by negative one.  We call the
term being maximized in \eqref{eq:elbo} the evidence lower bound (ELBO) because
it is a lower bound for the marginal distribution of $\Y$, $p(\Y)$, also called
the evidence in the machine learning literature \citep{vbreview}.

\subsection{Mean-Field Approximations} \label{sec:mfvb}

The tractability of the optimization problem depends on the family of densities
under consideration; the more complicated the family of densities the harder the
optimization problem will become. Additionally, once we find the optimal
density, we will need to calculate expectations and quantiles with respect to
it, which pushes us towards simpler approximations. One of the most popular
families to use in variational inference is to assume that the distribution of
subsets of the parameter vector, $\vtheta = (\vtheta^T_1, \dots,
\vtheta^T_P)^T$, are independent, where the subsets are chosen for computational
convenience. Hence, 
\begin{align}
\label{eq:vir:mfvb}
\mathcal{D} = \left\{q(\vtheta) : q(\vtheta) = \prod_{p = 1}^P 
q_p(\vtheta_p) \right\}. 
\end{align}
The class of densities in \eqref{eq:vir:mfvb} do not make an assumption
regarding the optimal distributions for each $\vtheta_p$ except for the fact
that they are independent. Additionally, the groups within $\vtheta$ can be
selected to with a particular structure in mind, with a focus on grouping
correlated parameters together. 

\subsubsection{Coordinate Ascent Variational Inference (CAVI)}

Coordinate ascent is the most popular approach for finding the optimal
distribution under the mean-field restrictions in \eqref{eq:vir:mfvb}. This
approach iteratively optimizes the ELBO with respect to each factorized density,
while holding all other constant. The resulting algorithm is dubbed coordinate
ascent variational interference (CAVI). The derivations for the optimal density
are given in various texts \cite{bishop2006pattern, vbreview, murphy2012} and we
state the result below for convenience. First, note that the ELBO with respect
to the factor, $q_p(\vtheta_p)$, can be written as:
\begin{align*}
\begin{split}
\text{ELBO}[q_p(\vtheta_p)] & = 
\E_{q(\vtheta)} \left[
\log p(\vtheta, \Y)
\right ] -
\sum_{j = 1}^P 	\E_{q_j(\vtheta_j)} \left[  \log q_j(\vtheta_j)	\right],
\\
& = 
\E_{q_p(\vtheta_p)} \left[ 
\E_{q_{r}(\vtheta_{r})} \left \{
\log p(\vtheta, \Y)
\right \}
\right ] -
\E_{q_p(\vtheta_p)} \left[ \log q_p(\vtheta_p) \right] + 
\text{const}, \\
& = - KL \left[
q_p(\vtheta_p), \ 
\exp \left\{ \E_{q_{r}(\vtheta_{r})} \left[
\log p(\vtheta, \Y)
\right] \right \}
\right] + \text{const},
\end{split}
\end{align*}
where we define $q_r(\vtheta_r)$ as the variational distribution for the rest of
the parameters: $q_{r}(\vtheta_r) = \prod_{j \neq p}^P q_{j}(\vtheta_j)$. Since
the $KL$ divergence between two probability distributions is always positive,
the negative KL divergence is maximized when the divergence between two
probability densities is equal to zero.  Hence the optimal variational density
for the $p^{th}$ factor is: 
\begin{align}
\label{eq:vir:cavi_optim}
q^*_p(\vtheta_p) & \propto \exp \left\{ \E_{q_{r}(\vtheta_{r})} \left[
\log p(\vtheta, \Y)
\right] \right \}, \\
& \propto \exp \left\{ \E_{q_{r}(\vtheta_{r})} \left[
\log p(\vtheta_p | \vtheta_r, \Y)
\right] \right \}.
\end{align}
Consequently, the optimal variational density for the $p^{th}$ coordinate is a
function of the full conditional distribution the parameter, a density that is
required calculation for a Gibbs sampler. 

It should be noted that, in most situations, the difficulty of calculating the
expectations in \eqref{eq:vir:cavi_optim} is a direct function of the simplicity
of the variational family in \eqref{eq:vir:mfvb}. Assuming that a larger number
of parameters independently factor in the posterior simplifies the calculations
of the expectations of, for example, the inner product of two vector parameters.

\subsection{Stochastic Variational Inference (SVI)}

If we further assume that the complete conditionals of the parameter given all
others are in the exponential family \eqref{eq:vir:exp-fam}, and require the
individual factors in the mean-field family \eqref{eq:vir:mfvb} to be in the
same exponential family, we can scale variational inference to large datasets
that do not fit in memory. This approach, termed Stochastic Variational
Inference (SVI) \citep{hoffman2013svi}, utilizes gradient based optimization to
maximize the ELBO instead of using the coordinate ascent algorithm presented in
Section \ref{sec:mfvb}. SVI compares favorable to CAVI in models that have local
variables; this means that for each data point $y_n$ there exists a latent
parameter $z_n$ that needs to be estimated. In such cases, both Gibbs sampling
and the CAVI algorithm process the entire dataset every iteration. 

For the rest of this section, we present SVI in the context of probit regression
with a normal (ridge) prior. This model has the hierarchy:
\begin{align}
\label{eq:probit_ridge}
\begin{split}
y_n & = \mathbb{I}(z_n > 0), \\
z_n & \sim \mathcal{N}(\x_n\Tra \b, 1), \\
\b & \sim \mathcal{N}(0, \lambda\Inv \I), \\
\lambda & \sim  \mathcal{G}(a_{\lambda}, b_{\lambda}).
\end{split}
\end{align}
In this situation, the observed data is $y_n$, and each $y_n$ has a
corresponding local variable, $z_n$, which allows the use of the probit link in
the calculation of $P(y_n = 1) = \Phi(\x_n\Tra \b)$ where $\Phi(\cdot)$ is the
CDF of the standard normal distribution. In a Gibbs sampler and a CAVI
algorithm, we would have to update each $z_n$ before we can update the parameter
vector $\b$, estimation of which is of primary interest. 

It is well known that the complete conditional distributions of the parameters
in \eqref{eq:probit_ridge} are given by:
\begin{align*}
\begin{split}
z_n | \cdot & \sim \begin{cases}
\mathcal{N}_{+} \left(\x_n\Tra \b, 1 \right) \text{if } y_n = 1 \\
\mathcal{N}_{-} \left(\x_n\Tra \b, 1 \right) \text{if } y_n = 0 \\
\end{cases}, \\
\b | \cdot & \sim \mathcal{N} \left(
(\XtX + \lambda \I)\Inv \X\Tra \z, \ 
(\XtX + \lambda \I)\Inv
\right), \\
\lambda | \cdot & \sim \mathcal{G} \left(
a_{\lambda} + \frac{P}{2}, 
\frac{1}{2} ||\b||^2_2 + b_{\lambda}
\right), 
\end{split}
\end{align*}
where, for example, $\b | \cdot$ denotes the distribution of $\b$ conditioned on
all other parameters in the model, $\X = (\x_1\Tra, \dots, \x_N\Tra)\Tra$, and
$\normal_{+}$ and $\normal_{-}$ are truncated normal distributions on $(0,
\infty)$ and $(-\infty, 0]$, respectively. Consequently, under this setup, we
would restrict the optimal variational distribution of $\b$ to be a normal
distribution, i.e. $q_{b}(\b) = \mathcal{N} \left( \vmu_b, \msigma_b \right)$.
Because the normal distribution is part of the exponential family, it can be
written in the form of \eqref{eq:vir:exp-fam}: 
\begin{align}
q(\b | \vdelta_b(\vtheta)) & = h(\b) \exp \left \{ 
\langle \vdelta_b (\vtheta) |  \t_b(\b) \rangle
- A \left[ \vdelta_b (\vtheta) \right]
\right \},
\end{align}
where,
\begin{align}
\t(\b) & = \begin{pmatrix}
\b \\
\b \b\Tra
\end{pmatrix} \text{ and } \vdelta_b(\vtheta) = \begin{pmatrix}
-\frac{1}{2} \msigma_b\Inv \vmu_b \\
-\frac{1}{2} \msigma_b \Inv
\end{pmatrix}.
\end{align}
Now, looking at the full conditional for $\b$, we can also put it into
exponential family form as:
\begin{align}
p(\b | \X, \z, \lambda) & = h(\b) \exp \left\{
\langle \vdelta_b (\X, \z, \lambda) |  \t_b(\b) \rangle
- A \left[ \vdelta_b (\X, \z, \lambda) \right]
\right\}. 
\end{align}
Instead of taking the gradient of the ELBO with respect to $\vdelta_b(\vtheta)$,
\citet{hoffman2013svi} utilize the natural gradient \citep{amari1998natural},
which accounts for the geometry of the parameter space.  Applying their
derivation for the natural gradient of the ELBO to our case
(\citet{hoffman2013svi} ; Equation 14), gives us the natural gradient with
respect to $\vdelta_b(\vtheta)$: 
\begin{align}
\label{eq:natural_grad}
\begin{split}
\nabla_{\vdelta_b(\vtheta)} ELBO & = \Eq[\vdelta_b(\X, \z, \lambda)] - 
\vdelta_b(\vtheta), \\
& = 
\Eq \left[ 
\begin{pmatrix}
\sum_{n = 1}^N z_n \x_n \\
\sum_{n = 1}^N \x_n \x_n\Tra + \lambda \I 
\end{pmatrix}
\right] - \vdelta_b(\vtheta), \\
& = 
\begin{pmatrix}
\sum_{n = 1}^N \Eq \left[ z_n \right] \x_n \\
\sum_{n = 1}^N \x_n \x_n\Tra + \Eq \left[ \lambda \right] \I 
\end{pmatrix} - \vdelta_b(\vtheta).
\end{split}
\end{align}
This expression can be used in a gradient descent algorithm where, at each
iteration, the natural parameters of the variational distribution,
$\vdelta_b(\vtheta)$, are updated using the following formula: 
\begin{align}
\label{eq:gradient_step}
\begin{split}
\vdelta_b(\vtheta)^{(t)} & = \vdelta_b(\vtheta)^{(t-1)} + 
\rho_t \left[ 
\Eq[\vdelta_b(\X, \z, \lambda)] - \vdelta_b(\vtheta)^{(t-1)}
\right], \\
& = (1 - \rho_t) \vdelta_b(\vtheta)^{(t-1)} + 
\rho_t \Eq[\vdelta_b(\X, \z, \lambda)] ,
\end{split}
\end{align}
where $\rho_t$ is a predetermined step size. 

Note that the gradient in \eqref{eq:natural_grad} requires the processing of all
data points due to the summation over $N$. Instead of calculating the gradient
with respect to the full data, we can calculate an approximation which is equal
to full gradient in expectation, and follow the iterative procedure in
\eqref{eq:gradient_step}. If the sequence of step sizes, $\rho_t$, meets the
conditions in \eqref{eq:rhot_cond}, $\vdelta_b(\vtheta)^{(t)}$ will converge to
a local optimum. 
\begin{align}
\label{eq:rhot_cond}
\begin{split}
\sum_t \rho_t = \infty \text{ and } \sum_t \rho_t^2 < \infty
\end{split}
\end{align}
A noisy estimate of the gradient can be calculated by using only one uniformly
sampled data point replicated $N$ times. This means that the summation in
\eqref{eq:natural_grad} would change to the value for one data point multiplied
by $N$; that is, the parameter update in \eqref{eq:gradient_step} becomes:
\begin{align}
\label{eq:sgradient_step}
\begin{split}
\vdelta_b(\vtheta)^{(t)} & = (1 - \rho_t) \vdelta_b(\vtheta)^{(t-1)} + 
\rho_t \left \{
N \Eq \left[ \vdelta_b(\x_n, z_n) \right]
\right \},
\end{split}
\end{align}
with 
\begin{align*}
N \Eq \left[ \vdelta_b(\x_n, z_n) \right] = 
\begin{pmatrix}
	N \Eq \left[ z_n \right] \x_n \\
	N \x_n \x_n\Tra + \Eq \left[ \lambda \right] \I
\end{pmatrix}.
\end{align*}
Finally, while the stochastic gradient updates in \eqref{eq:sgradient_step} are
only computed for one data point at a time, the methodology can be extended to
process S data points simultaneously. The only difference would be to average
the individual results from each data point. The update step becomes: 
\begin{align*}
\begin{split}
\vdelta_b(\vtheta)^{(t)} & = (1 - \rho_t) \vdelta_b(\vtheta)^{(t-1)} + 
\frac{\rho_t}{S} \left \{
N \sum_{s \in \mathcal{S}} \Eq \left[ \vdelta_b(\x_s, z_s) \right]
\right \},
\end{split}
\end{align*}
which implies that, 
\begin{align*}
N \sum_{s \in \mathcal{S}} \Eq \left[ \vdelta_b(\x_n, z_n) \right] = \begin{pmatrix}
	N \X_s\Tra \Eq \left[ \z_s \right] \\
	N \X_s\Tra \X_s + S \Eq \left[ \lambda \right] \I \\
\end{pmatrix},
\end{align*}
where $\mathcal{S}$ is the set of sub-sampled data points, $S = |\mathcal{S}|$,
and $\X_s$ and $\z_s$ are $\X$ and $\z$ restricted to the corresponding indices
in $\mathcal{S}$. 

\section{Implementation and Usage} \label{sec:vir:usage}

The package \pkg{vir} contains implementations of the CAVI and SVI algorithms
for univariate and multivariate regression models with the normal and probit
link.  For the univariate linear models, we derive and implement a Gibbs
sampler, and the CAVI and SVI algorithms for the ridge (normal), LASSO
(double-exponential) \citep{park2008lasso}, and horseshoe
\citep{carvalho2010horseshoe, makalic2015simple} priors, while the multivariate
linear models are implemented with non-informative priors for the regression
coefficients and a factor model for the covariance structure. Each function is
implemented in \proglang{C++} and leverages the \pkg{Rcpp} \citep{rcpp} and
\pkg{RcppEigen} \citep{bates2013fast} packages to optimize performance. 

The functions in the package are named according to the link function, \code{lm}
for normal linear models and \code{probit} for binary regression, followed by
the shrinkage prior type: \code{ridge}, \code{lasso}, \code{hs}, \code{uninf};
and then the algorithm: \code{gibbs}, \code{cavi}, or \code{svi}. Therefore, if
an analyst wishes to use the svi algorithm with a linear model and horseshoe
prior, they can call the function \code{lm\_hs\_svi} to analyze the data. Table
\ref{tab:implementation} contains a summary of the names for the functions in
the package.  All functions take, as arguments, a matrix of predictors, $\X$ and
a vector or matrix of responses, $\y$ or $\Y$. 

The variational algorithms utilize two assumptions regarding the regression
coefficients. The first assumes that each regression coefficient is correlated,
and sets the optimal distribution of the parameter to be a multivariate normal,
that is: $q(\b) = \mathcal{N} \left(\vmu_b, \msigma_b \right)$. Alternatively,
we also implement a version of the algorithm which assumes that all regression
coefficients are independent, i.e., $q(\b) = \prod_{p = 1}^P q(b_p)$, where each
component, $q(b_p)$ is a univariate normal distribution. The use of these two
options is problem dependent; assuming the full correlation structure may lead
to superior variance estimates at the expense of slower computation time. 

\begin{table}[t]
	\centering
		\caption{List of functions implemented in the \pkg{vir} package. The 
		functions 
		have the format \code{model\_prior\_algorithm}. Consequently, if an analyst 
		wishes to use a Gibbs sampler to fit a multivariate linear model with a 
		non-informative prior then they could call the function 
		\fct{mv\_lm\_uninf\_gibbs}. If they wished to fit a univariate probit model 
		with a 
		horseshoe prior using CAVI, they would call the function 
		\fct{probit\_hs\_cavi}. 
		Each function is implemented in \proglang{C++} using the \pkg{RcppEigen} 
		package.} \label{tab:implementation} 
	\begin{tabular}{Sc|Sc|Sc}
		\hline
		\multicolumn{1}{Sc|}{\textbf{Model}}

		               & 
		\multicolumn{1}{c|}{\textbf{Priors}}

		   &
		 \multicolumn{1}{c}{\textbf{Algorithms}} \\ \hline
		\bigcell{c}{Normal: \code{lm} \\ Probit: 
		\code{probit}}                                                  & \bigcell{c}{Ridge: 
		\code{ridge} \\ LASSO: \code{lasso} \\ Horseshoe: \code{hs}} & 
		\multirow{4}{*}{\bigcell{c}{Gibbs: \code{gibbs} \\ CAVI: \code{cavi} \\ SVI: 
		\code{svi}}}  \\
		\cline{1-2} 
		\bigcell{c}{Multivariate Normal: \code{mv\_lm} \\ Multivariate Probit: 
		\code{mv\_probit}} & \bigcell{c}{Non-informative: \code{uninf}}            \\   
		\hline
	\end{tabular}
\end{table}

Below we provide an example use case, analyzing a simulated dataset with a
univariate normal linear model \eqref{eq:lm_sim}. We first demonstrate the use
of CAVI with a normal prior. As described earlier, the function to be called in
this situation is \fct{lm\_ridge\_cavi}; the documentation for which can be seen
by executing \code{?lm\_ridge\_cavi} in the \proglang{R} console. We install the
package from GitHub and start by simulating a small dataset and fitting the
model.
\begin{CodeChunk}
\begin{CodeInput}
> devtools::install_github("suchitm/vir")
> library(vir)
> set.seed(42)
> X = matrix(nrow = 100, ncol = 5, rnorm(5 * 100))
> colnames(X) = paste0("X", 1:5)
> b = rnorm(5)
> y = rnorm(1) + X %*% b + rnorm(100)
> ridge_cavi_fit = lm_ridge_cavi(y, X, n_iter = 100, rel_tol = 0.0001)
\end{CodeInput}
\end{CodeChunk}
Each of the algorithms output a nested list. The first layer contains an element
for each of the parameters and the ELBO, while the second lists the form of the
optimal distribution (normal, gamma, etc.), and the corresponding optimal
parameters. Since the normal ridge model has four parameters, the first level of
the list has the names: 
\begin{CodeChunk}
\begin{CodeInput}
> names(ridge_cavi_fit)
\end{CodeInput}
\begin{CodeOutput}
[1] "b0"     "b"      "tau"    "lambda" "elbo"  
\end{CodeOutput}
\end{CodeChunk} 
where \code{b0} is the intercept term, \code{b} is a vector of regression
coefficients, \code{tau} is the error precision, and \code{lambda} is the prior
precision. The optimal distribution of the regression coefficients, $\b$, is a
multivariate normal with mean, $\vmu$, and covariance matrix, $\msigma$, the
names for that parameter element contain the distribution type and corresponding
parameters. 
\begin{CodeChunk}
\begin{CodeInput}
> names(ridge_cavi_fit$b)
\end{CodeInput}
\begin{CodeOutput}
[1] "dist"      "mu"        "sigma_mat"
\end{CodeOutput}
\end{CodeChunk}
\begin{CodeChunk}
\begin{CodeInput}
> ridge_cavi_fit$b$dist
> ridge_cavi_fit$b$mu
> round(ridge_cavi_fit$b$sigma_mat, 4)
\end{CodeInput}
\begin{CodeOutput}
[1] "multivariate normal"
[1]  0.87668826  0.92639922  0.08866204  0.10254637 -0.75907903
        [,1]    [,2]    [,3]    [,4]    [,5]
[1,]  0.0110 -0.0006  0.0017 -0.0006 -0.0010
[2,] -0.0006  0.0144 -0.0011 -0.0005  0.0017
[3,]  0.0017 -0.0011  0.0115  0.0007 -0.0010
[4,] -0.0006 -0.0005  0.0007  0.0156 -0.0025
[5,] -0.0010  0.0017 -0.0010 -0.0025  0.0118
\end{CodeOutput}
\end{CodeChunk}
These parameter estimates can be summarized via the mean and credible 
intervals using \fct{summary\_vi}. 
\begin{CodeChunk}
\begin{CodeInput}
> summary_vi(ridge_cavi_fit, level = 0.95, coef_names = colnames(X))
\end{CodeInput}
\begin{CodeOutput}
             Estimate      Lower       Upper
Intercept -0.17491248 -0.3901667  0.04034176
X1         0.87668826  0.6711879  1.08218858
X2         0.92639922  0.6909048  1.16189365
X3         0.08866204 -0.1218641  0.29918818
X4         0.10254637 -0.1424468  0.34753954
X5        -0.75907903 -0.9717937 -0.54636432
\end{CodeOutput}
\end{CodeChunk}
Using the model fit, one can generate estimates and corresponding credible 
intervals for new data by utilizing the \fct{predict\_lm\_vi} 
function. 
\begin{CodeChunk}
\begin{CodeInput}
> X_test = matrix(nrow = 5, ncol = 5, rnorm(25))
> predict_lm_vi(ridge_cavi_fit, X_test)
\end{CodeInput}
\begin{CodeOutput}
$estimate
[1] -1.0235563  2.3194540 -0.5776915 -0.6762113 -1.2336189

$ci
           [,1]      [,2]
[1,] -3.1989515 1.1518389
[2,]  0.1016273 4.5372806
[3,] -2.7163884 1.5610055
[4,] -2.8682399 1.5158172
[5,] -3.4223356 0.9550978
\end{CodeOutput}
\end{CodeChunk}

\subsection{SVI}

The are three material differences between the CAVI and SVI implementations,
with the first being the stopping criteria. The CAVI algorithms use a relative
tolerance criteria (\code{rel\_tol}) comparing the ELBO at the current iteration
with the ELBO five iterations in the past. On the other hand, the SVI algorithms
do not terminate based on a stopping criteria, but will instead run until the
maximum number of iterations (\code{n\_iter}) is reached. For each sub-sample of
the data, a noisy estimate for the ELBO is calculated and stored, which can then
be used to assess convergence of the algorithms. 

Second, the step sizes, $\rho_t$, in stochastic gradient algorithms have to
follow the schedule outlined in \eqref{eq:rhot_cond}. We implement two ways of
meeting the necessary conditions. The default approach allows setting a constant
step size via the parameter \code{const\_rhot}, which keeps the step size
constant for each iteration of the algorithm. Alternatively, we also allow the
user to specify a schedule based on the equation $\rho_t = (t +
\omega)^{-\kappa}$, where $\omega \geq 0$ is called a delay while $\kappa \in
(0.5,  1]$ is called the forgetting rate \citep{hoffman2013svi}. Finally, SVI
algorithms utilize sub-samples of the data, and the size of these samples has to
be set via the \code{batch\_size} parameter.  Below we show how the
corresponding SVI algorithm for normal linear regression with a normal prior can
be implemented in our package; in this case we set a constant learning rate of
$0.1$ and use a sub-sample of ten data points for each iteration. After fitting
the model, all other operations regarding parameter summaries and predictions
are identical to the CAVI exposition above. 
\begin{CodeChunk}
\begin{CodeInput}
> ridge_svi_fit = lm_ridge_svi(
+   y, X, n_iter = 1000, verbose = FALSE, batch_size = 10, const_rhot = 0.1
+ )
\end{CodeInput}
\end{CodeChunk}

It should be noted that their exists an interaction between the batch size and
the step size for any stochastic gradient descent algorithm. For our numerical
experiments in Section \ref{sec:vir:simulations}, we use a constant learning
rate of 0.01, which seemed to work well for our purposes. Consequently, this is
the default rate for our algorithms. The choice of adaptive learning rates is an
open area of research in the stochastic optimization literature
\citep{nopeskylr, kingma2014adam} and we leave their implementation to later
iterations of our package. 

\section{Simulations} \label{sec:vir:simulations}

In this section we provide a simulation study comparing the ridge and LASSO
penalties implemented in \pkg{glmnet}, to a Gibbs sampler, and the CAVI and SVI
algorithms for linear and probit regression with ridge, LASSO, and horseshoe
priors. For each set of simulations we generate $D = 50$ datasets for our
comparisons.

\subsection{Linear Regression} \label{sec:lm_sims}

In this section, we simulate from the model 
\begin{align} \label{eq:lm_sim}
	\y = \boldsymbol{1} b_0 + \X \b + \e; \ 
		\e \sim \normal \left( \boldsymbol{0}, \sigma^2 \I \right),
\end{align} 
where $b_0 \sim \normal(0, 1)$, the $n^{th}$ row of $\X$, $\x_n \sim
\normal(\zerovec, \M{V})$, $\M{V}_{j, j'} = cov(x_j, x_{j'}) = 0.5^{|j - j'|}$,
$\b \sim \normal(\zerovec, \I)$, and $\sigma^2$ is chosen to set the
signal-to-noise ratio in the data to one.  We hold the number of predictors $P$
constant at $75$, and vary the size of the dataset, $N$.  We set 80\% of the
values in $\b$ to zero, which motivates our comparison of the shrinkage priors
below. 

\begin{table}[t]
	\caption{Simulation results for the normal linear model.}	
	\label{tab:lm_sims}
\begin{center}
\begin{tabular}{lccccccccccc}
\hline\hline
\multicolumn{1}{l}{\bfseries }&\multicolumn{3}{c}{\bfseries N = 100}&\multicolumn{1}{c}{\bfseries }&\multicolumn{3}{c}{\bfseries N = 1000}&\multicolumn{1}{c}{\bfseries }&\multicolumn{3}{c}{\bfseries N = 5000}\tabularnewline
\cline{2-4} \cline{6-8} \cline{10-12}
\multicolumn{1}{l}{}&\multicolumn{1}{c}{MSE}&\multicolumn{1}{c}{Cov.}&\multicolumn{1}{c}{MSPE}&\multicolumn{1}{c}{}&\multicolumn{1}{c}{MSE}&\multicolumn{1}{c}{Cov.}&\multicolumn{1}{c}{MSPE}&\multicolumn{1}{c}{}&\multicolumn{1}{c}{MSE}&\multicolumn{1}{c}{Cov.}&\multicolumn{1}{c}{MSPE}\tabularnewline
\hline
{\bfseries Ridge}&&&&&&&&&&&\tabularnewline
~~GLM-1SE&0.160&-&0.917&&0.038&-&0.732&&0.012&-&0.708\tabularnewline
~~GLM-Min&0.160&-&0.913&&0.039&-&0.732&&0.012&-&0.707\tabularnewline
~~Gibbs-Corr&0.125&0.941&0.856&&0.024&0.948&0.718&&0.005&0.951&0.700\tabularnewline
~~CAVI-Corr&0.135&0.946&0.868&&0.024&0.949&0.718&&0.005&0.952&0.700\tabularnewline
~~CAVI-Indep&0.122&0.913&0.853&&0.024&0.871&0.718&&0.005&0.871&0.700\tabularnewline
~~SVI-Corr&0.135&0.946&0.866&&0.026&0.943&0.721&&0.008&0.893&0.703\tabularnewline
~~SVI-Indep&0.122&0.913&0.854&&0.025&0.868&0.720&&0.006&0.819&0.704\tabularnewline
\hline
{\bfseries LASSO}&&&&&&&&&&&\tabularnewline
~~GLM-1SE&0.129&-&0.859&&0.029&-&0.721&&0.010&-&0.705\tabularnewline
~~GLM-Min&0.131&-&0.864&&0.029&-&0.721&&0.010&-&0.705\tabularnewline
~~Gibbs-Corr&0.120&0.968&0.852&&0.017&0.966&0.713&&0.004&0.963&0.700\tabularnewline
~~CAVI-Corr&0.133&0.937&0.863&&0.018&0.958&0.713&&0.004&0.959&0.700\tabularnewline
~~CAVI-Indep&0.106&0.901&0.833&&0.017&0.911&0.712&&0.004&0.897&0.700\tabularnewline
~~SVI-Corr&0.129&0.938&0.859&&0.019&0.952&0.714&&0.006&0.908&0.701\tabularnewline
~~SVI-Indep&0.091&0.917&0.814&&0.014&0.930&0.709&&0.005&0.877&0.701\tabularnewline
\hline
{\bfseries HS}&&&&&&&&&&&\tabularnewline
~~Gibbs-Corr&0.097&0.945&0.809&&0.010&0.975&0.705&&0.002&0.983&0.698\tabularnewline
~~CAVI-Corr&0.113&0.909&0.834&&0.011&0.956&0.706&&0.002&0.967&0.698\tabularnewline
~~CAVI-Indep&0.099&0.888&0.813&&0.010&0.931&0.706&&0.002&0.941&0.698\tabularnewline
~~SVI-Corr&0.110&0.908&0.831&&0.011&0.954&0.706&&0.002&0.961&0.698\tabularnewline
~~SVI-Indep&0.099&0.888&0.812&&0.010&0.933&0.705&&0.002&0.937&0.698\tabularnewline
\hline
\end{tabular}\end{center}

\end{table}

We compare the mean squared error for the coefficients (MSE), coverage for 95\%
credible intervals (Cov.), and the relative mean squared prediction error (MSPE)
for predictions on a test set of size 500 simulated from the model in
\eqref{eq:lm_sim}. The formulas to calculate our metrics are below:
\begin{align*} 
\begin{split}
\text{MSE} & = \frac{1}{DP} \sum_{d = 1}^{D} \sum_{p = 1}^P 
(b_p - \hat{b}_p)^2, \\
\text{MPSE} & = \sqrt{\frac{||\X \b - \X \hat{\b}||^2_2}{ ||\X \b||^2_2}}, \\
\text{COV} & = \frac{1}{DP} \sum_{d = 1}^D \sum_{p = 1}^P
\mathbb{I}(l_p \leq b_p \leq u_p),
\end{split}
\end{align*}
where $b_p$ is the true coefficient, $\hat{b}_p$ is the corresponding estimated
value, $\X$ is the predictor matrix for the test set, and $(l_{p}, u_p)$ is the
credible set for the $p^{th}$ predictor. 

The results for our simulations are given in Table \ref{tab:lm_sims}. As can be
seen, all implementations of the variational algorithms are competitive with
\pkg{glmnet} and the Gibbs sampler in regards to MSE and MSPE. The variational
algorithms for linear regression with the horseshoe prior provide superior
parameter estimation with respect to the LASSO or ridge priors, along with
outperforming the ridge and LASSO \pkg{glmnet} implementations. 

With regards to coverage, the variational algorithms which assume posterior
independence between the regression parameters have the lowest coverage among
the Bayesian approaches. However, the coverage improves as $N$ increases for the
horseshoe priors, which has approximately 94\% coverage for the 95\% credible
sets measured. On the other hand, the variational algorithms which estimate a
correlated structure among the parameters seem to retain coverage similar to
that of the Gibbs samplers while being estimated at a fraction of the
computational cost.  Therefore, while, in general variational algorithms are
known to underestimate the posterior variance of the model parameters, this does
not seem to be an issue in the simulation settings considered. Additionally, the
SVI and CAVI algorithms perform comparably in terms of point and variance
estimation, providing confidence for using SVI approaches in the context of a
linear regression where the data does not fit in memory. 

\subsection{Probit Regression} \label{sec:probit_sims}

We compare the probit regression algorithms in the package using a similar
design as in Section \ref{sec:lm_sims} with the modification that $z_n  = b_0 +
\x_n\Tra \b$ and $P(y_n = 1) = \Phi(z_n)$ for the probit link or $P(y_n = 1) =
\text{expit}(z_n)$ for the logit link. Because \pkg{glmnet} implements the
logistic link instead of the probit, we simulate from both the logistic and
probit models and compare the algorithms using only one design with $N = 500$,
$P = 50$, and set 40 of the parameters to zero.

\begin{table}[t]
	\caption{Simulation results for binary regression with shrinkage priors.}	
	\label{tab:probit_sims}
\begin{center}
\begin{tabular}{lcccccccc}
\hline\hline
\multicolumn{1}{l}{\bfseries }&\multicolumn{3}{c}{\bfseries Logit}&\multicolumn{1}{c}{\bfseries }&\multicolumn{4}{c}{\bfseries Probit}\tabularnewline
\cline{2-4} \cline{6-9}
\multicolumn{1}{l}{}&\multicolumn{1}{c}{MSE}&\multicolumn{1}{c}{RAND}&\multicolumn{1}{c}{AUC-PR}&\multicolumn{1}{c}{}&\multicolumn{1}{c}{MSE}&\multicolumn{1}{c}{Coverage}&\multicolumn{1}{c}{RAND}&\multicolumn{1}{c}{AUC-PR}\tabularnewline
\hline
{\bfseries Ridge}&&&&&&&&\tabularnewline
~~GLM-1SE&0.079&0.673&0.881&&-&-&0.673&0.924\tabularnewline
~~GLM-Min&0.174&0.673&0.885&&-&-&0.673&0.922\tabularnewline
~~Gibbs-Corr&-&0.852&0.885&&0.023&0.951&0.850&0.930\tabularnewline
~~CAVI-Corr&-&0.772&0.886&&0.034&0.700&0.706&0.929\tabularnewline
~~CAVI-Indep&-&0.671&0.886&&0.035&0.588&0.607&0.929\tabularnewline
~~SVI-Corr&-&0.765&0.885&&0.031&0.692&0.692&0.929\tabularnewline
~~SVI-Indep&-&0.660&0.885&&0.032&0.568&0.588&0.929\tabularnewline
\hline
{\bfseries LASSO}&&&&&&&&\tabularnewline
~~GLM-1SE&0.180&0.819&0.894&&-&-&0.794&0.934\tabularnewline
~~GLM-Min&0.174&0.621&0.898&&-&-&0.591&0.938\tabularnewline
~~Gibbs-Corr&-&0.866&0.892&&0.024&0.970&0.863&0.936\tabularnewline
~~CAVI-Corr&-&0.811&0.895&&0.016&0.782&0.790&0.939\tabularnewline
~~CAVI-Indep&-&0.779&0.895&&0.019&0.740&0.763&0.939\tabularnewline
~~SVI-Corr&-&0.810&0.894&&0.017&0.782&0.790&0.939\tabularnewline
~~SVI-Indep&-&0.846&0.898&&0.019&0.779&0.829&0.942\tabularnewline
\hline
{\bfseries HS}&&&&&&&&\tabularnewline
~~Gibbs-Corr&-&0.845&0.900&&0.012&0.981&0.875&0.943\tabularnewline
~~CAVI-Corr&-&0.871&0.901&&0.010&0.856&0.869&0.944\tabularnewline
~~CAVI-Indep&-&0.869&0.901&&0.010&0.845&0.873&0.944\tabularnewline
~~SVI-Corr&-&0.871&0.901&&0.009&0.857&0.868&0.944\tabularnewline
~~SVI-Indep&-&0.868&0.901&&0.009&0.845&0.870&0.944\tabularnewline
\hline
\end{tabular}\end{center}

\end{table}

We calculate the mean squared error for the parameter estimates for each of the
algorithms when the data generating process is the same as the one assumed by
the model (logistic for \pkg{glmnet} and probit for \pkg{vir}), and compare the
coverage of the parameter estimates in the probit case for the Gibbs sampler and
the variational algorithms. We also compare the variable selection quality of
the respective algorithms by using the RAND index \cite{rand1971objective},
thinking of the variable selection problem as one of calculating the
dissimilarities between two cluster indicator vectors. Finally, we compare the
predictive capacity of our models by using the area under the precision recall
curve (AUC-PR). The results for our simulation are given in Table
\ref{tab:probit_sims}. 

In both simulation settings, the Gibbs samplers remain the most consistent in
terms of the RAND index. The CAVI and SVI algorithms with the horseshoe prior
perform comparably to the Gibbs samplers but the variational algorithms with the
ridge prior perform meaningfully worse, while the ones with the LASSO perform
only slightly worse. All algorithms perform similarly in terms of AUC-PR, with
the horseshoe prior being slightly better than the others.

In terms of parameter estimation in the probit simulation case, we see that the
estimation quality of the variational algorithms is comparable to the Gibbs
samplers for the LASSO and horseshoe priors, while being worse for the ridge
prior. As with the linear regression simulations, the performance of the SVI and
CAVI algorithms is similar, giving us confidence in their use for large scale
regression problems. 

Unfortunately, when it comes to coverage, the variational algorithms
meaningfully under perform the corresponding Gibbs samplers, likely due to the
mean-field assumption that the latent variables for each data point are
uncorrelated with the regression coefficients.  \citet{fasano2019scalable} show
that a mean-field approach for probit regression causes the regression estimates
to be shrunk towards zero, and propose a variational algorithm in the spirit of
the Gibbs sampler developed by \citet{holmes2006bayesian} to remedy the issue.
However, their proposed algorithm requires the prior for the regression
coefficients be fixed a-priori, and their updates for the latent variables
require the whole data set. Consequently, their algorithms cannot be readily
extended to the adaptive shrinkage priors considered in this article, and their
CAVI algorithm cannot be easily extended to utilize stochastic optimization.
Therefore, we recommend that the estimated posterior variance of the regression
coefficients be viewed with skepticism for the variational algorithms and note
that if one desires accurate uncertainty quantification in this situation, one
may, ironically, be able to utilize the bootstrap \citep{chen2018use}.

\section{Computational Performance} \label{sec:vir:timing}

A motivating factor for the development of our package was the computational
efficiency provided by variational methods relative to MCMC; we aimed to scale
linear regression with shrinkage priors to large datasets.  While it is known
that variational algorithms outperform their MCMC counterparts computationally,
in this section we provide timing comparisons of our implementations for linear
and binary regression to those in the package \pkg{glmnet}. Before we proceed to
a discussion of the relative timing results for each model, it should be noted
that the variational algorithms provide relatively accurate uncertainty
quantification for normal linear models while \pkg{glmnet} implementations
provide only a point estimate; in fact, frequentist uncertainty quantification
for $L1$ penalties is an open area of research  \citep{kyung2010penalized}.
Additionally, \pkg{glmnet} implements a path algorithm for estimating the
regression coefficients at various values of the tuning parameter, which is not
easily extendable to utilize stochastic optimization. Therefore, to obtain point
estimates for stochastic optimization based approaches for the LASSO, one would
have to perform cross-validation to find the optimal value of the tuning
parameter, an issue that does not exist for variational algorithms. 

We conducted two sets of timing experiments. First, we fixed $N$ at 1000 and
varied $P$ from 100 to 800 (with 80\% set to zero) to estimate how our methods
scale with the number of predictors. Second, we fixed $P$ at 100 and varied $N$
from 1000 to 50000 to estimate how our methods scale with dataset size. We
simulated five different datasets and calculated the time to convergence for
$\fct{cv.glmnet}$ and the CAVI algorithms with a relative tolerance of $0.0001$.
We ran the SVI algorithms for 15,000 iterations in the increasing $N$ case for
the binary regressions, since the primary reason to use SVI is its applicability
in large $N$ settings where there are a large number of local variables that
need to be updated. 

\begin{figure}
	\centering
	\includegraphics[scale=0.5]{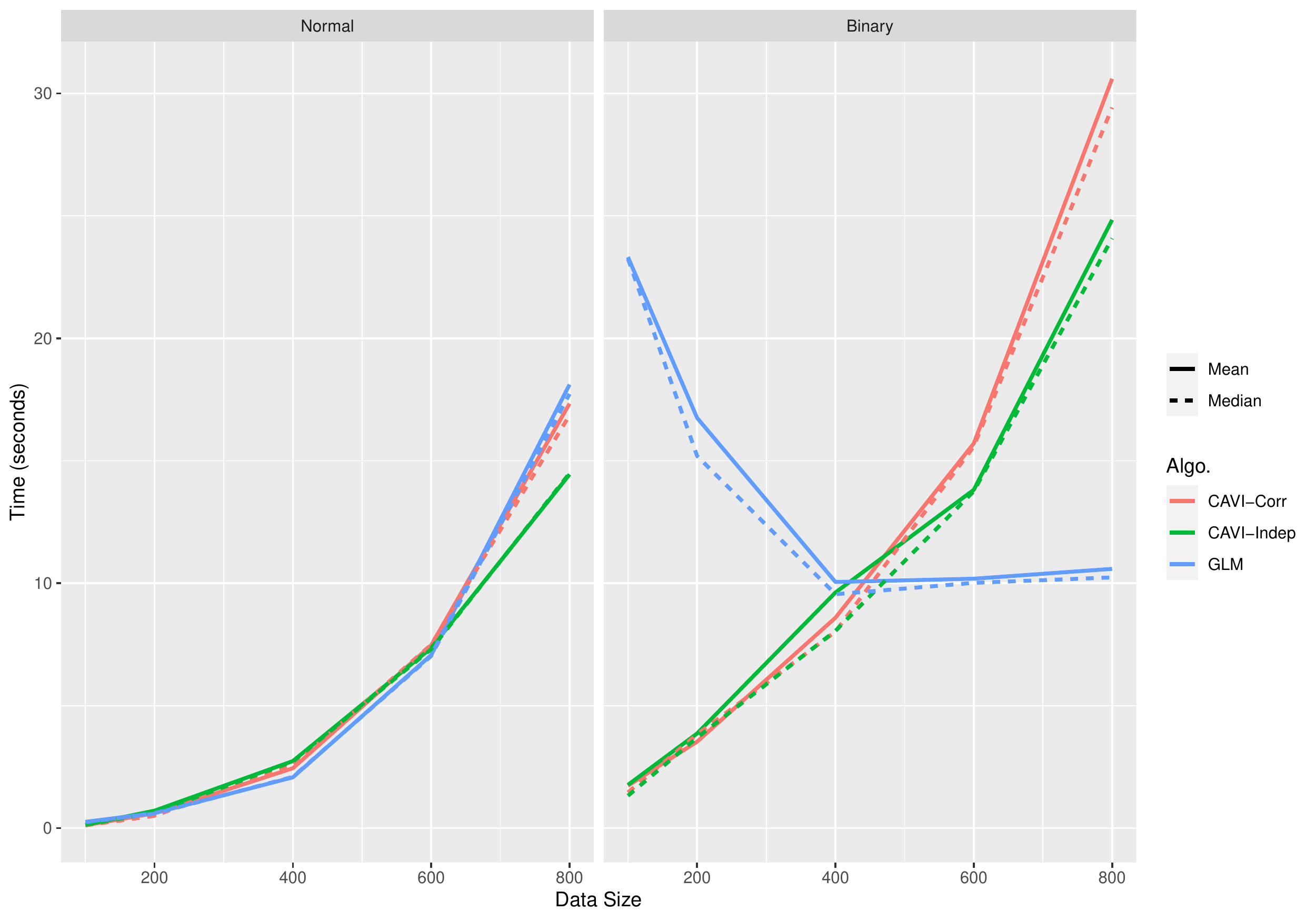}
	\caption{Timing results as the number of predictors is varied with 
		sample size fixed.}	\label{fig:timing_n_fix}
\end{figure}

\begin{figure}
	\centering
	\includegraphics[scale=0.5]{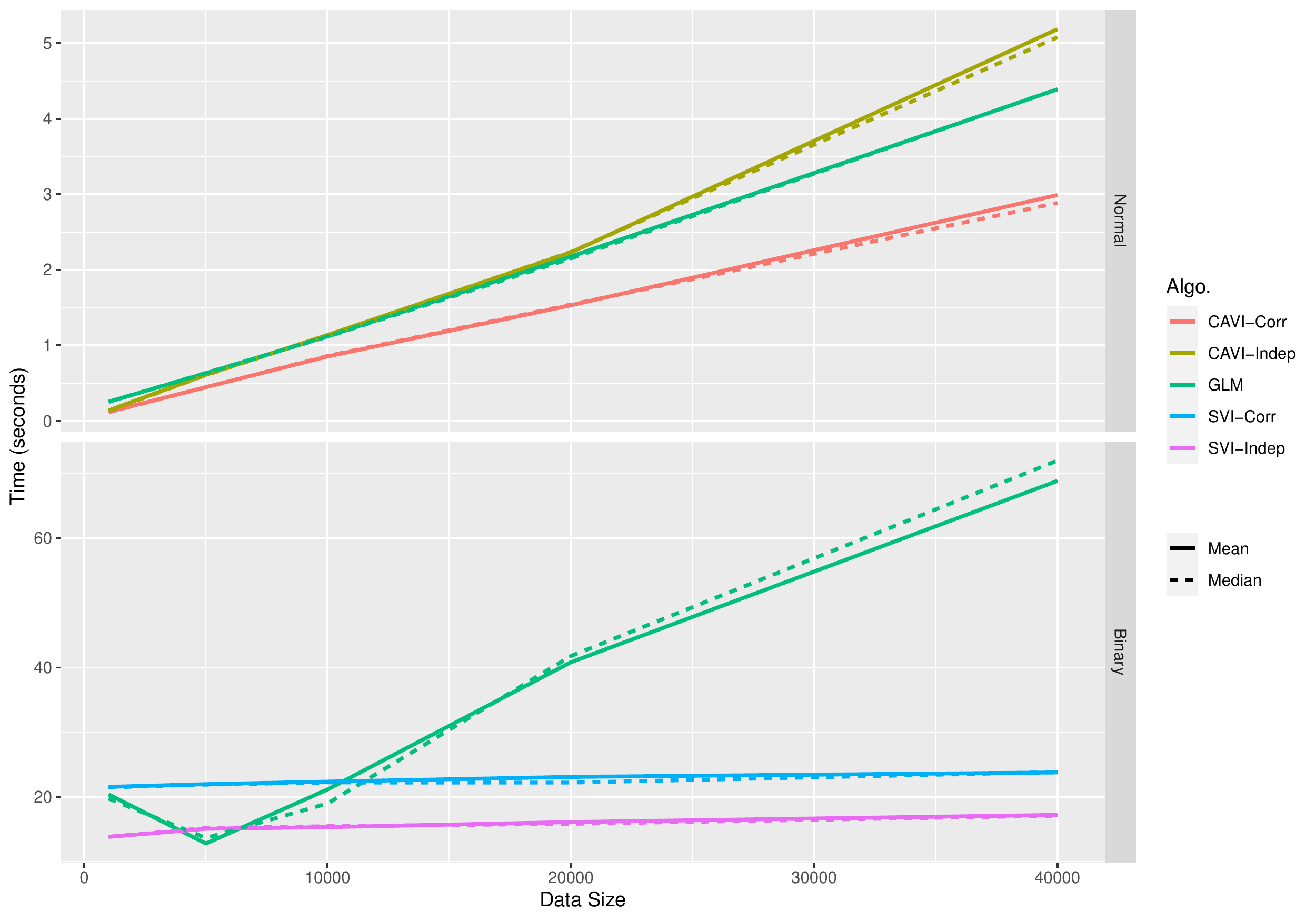}
	\caption{Timing results when the sample size is varied with the number 
		of predictors fixed at 100.}	\label{fig:timing_p_fix}
\end{figure}

Figure \ref{fig:timing_n_fix} shows our results for the timing experiments when
$N$ is fixed and $P$ is varied. For the linear models, we see that the time to
convergence increases for each model as the number of predictors is increased.
Interestingly, we see that, for the settings considered, our CAVI
implementations for normal linear models converge faster than the path algorithm
implemented in \fct{cv.glmnet}. It can also be seen that assuming an
independence structure among the regression coefficients can lead to
computational benefits as the number of predictors increases. In the case of the
probit regression models, we see, as expected, that the time to convergence for
the CAVI algorithms increases with the number of predictors. However,
perplexingly, the \fct{cv.glmnet} implementation does not seem to increase in
computational complexity as the number of predictors increases.

Figure \ref{fig:timing_p_fix} shows the time to convergence when $P$ is fixed
and $N$ is varied. We see that in the normal linear model case, our CAVI-Corr
implementations continue to outperform \fct{cv.glmnet} as the sample size
increases. It should be noted that the CAVI-Indep implementation is slower than
CAVI-Corr because a large error term needs to be calculated for each predictor
in each iteration of the algorithm. In the binary regression cases, the
CAVI implementations were substantially slower the \pkg{glmnet} implementations 
and were omitted from the plot.  However, since Bayesian binary regression is an
appropriate use case for the SVI algorithm even when the data fits in memory, we
see that the use of such algorithms for a fixed batch size and number of
iterations done not lead to an increase in computation time as the size of the
dataset increases. 

\section{Discussion} \label{sec:vir:discussion}

In this chapter we proposed a new \proglang{R} package, \pkg{vir}, for
computationally efficient Bayesian linear regression with shrinkage priors.  We
compared its performance to \pkg{glmnet}, which is one of the most widely used
tools for quickly performing variable selection and prediction in linear
regression. 

We conducted a simulation study which showed that variational algorithms can be
relied upon for variable selection and their performance is comparable to MCMC
based approaches for parameter estimation and uncertainty quantification in
normal linear models. Hence, we have provided a new tool in the Bayesian toolbox
for quickly exploring a large number of models before doing a final MCMC based
analysis. Second, our timing comparisons showed situations in which our package
outperforms state of the art approaches in \pkg{glmnet}, while at the same time
providing approximate uncertainty quantification for the coefficients.
Additionally, in the normal linear model case, our approach does not require the
use of the bootstrap to do uncertainty quantification, which creates substantial
computational overhead for frequentist penalization-based approaches to linear
regression. 

Future work will focus on increasing the number of variational algorithms
implemented in our package. At present, we focused on linear and binary
regression, and will extend the algorithms to count and survival data. We also
aim to add additional methods for clustering with Dirichlet Processes and
non-parametric regression models.  Finally, while variational algorithms are one
way to approximate a posterior distribution for large datasets, our long term
goal is to create a package which solves common problems using a variety of
methodology for big data analysis in Bayesian models (stochastic gradient MCMC,
GPU acceleration, etc.). 	

\bibliography{references}
\bibliographystyle{plainnat}

\end{document}